\documentclass[conference]{IEEEtran}
\usepackage{cite}
\usepackage{bm}
\usepackage{float}

\ifCLASSINFOpdf
\usepackage[pdftex]{graphicx}
\else
\fi
\usepackage{amsthm,amsmath,amssymb}
\usepackage{mathrsfs}
\usepackage{array}
\begin{document}
%
\title{Influence of User Emotion on Information Propagation with Public Sentiment in the Chinese Sina-microblog}

\author{\IEEEauthorblockN{Fulian Yin, Xiaojian Zhang, Nan Song, Xinyu Xia, Jiahui Lv}
\IEEEauthorblockA{College of Information and\\Communication Engineering,\\
Communication University of China\\
Beijing,China 100024\\
Email: yinfulian@cuc.edu.cn}
\and
\IEEEauthorblockN{Jianhong Wu}
\IEEEauthorblockA{Fields-CQAM Laboratory of Mathematics\\for Public Health,\\
Laboratory for Industrial and Applied Mathematics,\\
York University, Toronto, M3J1P3, Canada\\
Email: wujhhida@hotmail.com}}


%


\maketitle

\begin{abstract}
Social networks are flooded with different pieces of emotional information, the propagation of which helps to shape the development of public sentiment. To help designing effective communication strategies during the entire development of an event, we propose an emotion-based susceptible-forwarding-immune (E-SFI) propagation dynamic model, that takes into account of the categories of emotions into positive, neutral and negative and the emotional choices of user communities, to investigate the information propagation process that leads to public sentiment.
Our Model-based analytic and numerical analyses show that three types of forwarding probabilities involved in our E-SFI model are in accordance with the actual accident situation, and our sensitivity analyses describe important factors that affect the emotional choices of user communities in support for decision strategies to guide the public sentiment.
\end{abstract}

\def\IEEEkeywordsname{Keywords}

%
\IEEEpeerreviewmaketitle

\section{Introduction}
In the New Media Era, messages posted, forwarded and/or commented over the Internet carry not only information, ideas and insights, but also emotions. This rapidly evolving information and emotion propagation shapes the public sentiment, and leads to consistent accumulation of  comprehensive reflections of opinion polls. Depending on the subject issue of an event, during the development of the event and the relevant information propagation, diversified public sentiments are constantly exchanged and reshaped. In an emergency situation such as the early stage of the emerging COVID-19 pandemic, it is critical to ensure the public sentiment to reach a desired equilibrium state quickly to guide the implementation of appropriate public health measures in the user community and to achieve emotional relief. Challenges arise to reshape the public sentiment as various emotions are expressed and magnified in the social media, and call for some communication strategies to optimize the way public messaging in a highly dynamic process.

The propagation process of a piece of information and some potential way to  promote the emergence of a desired public sentiment can be described as follows. An original post owner released a piece of information and expressed his emotion, which could be positive, neutral or negative, depending on the content of the message. After reading the information containing the emotion of the original post owner, some users will forward the information with their own emotion, either positive, neutral or negative. Subsequently, a new portion of users exposed the information from those who read the original message with their own emotions, and the propagation chain of information with emotions continue until it reaches a certain equilibrium. As users who forward the information often express their emotions, directly or indirectly, the public sentiment during this propagation chain in a public hot event adapts and an equilibrium emerges. It is an important research topic, and our main objective of this study, to develop an appropriate model and analysis for this information and public sentiment propagation to identify effective communication strategies to guide the emergency of public sentiment towards a coherent collective response to emergency management.

The development of dynamic models to understand mechanism of information dissemination has always been a topic of interest to challenge modellers. In particular, infectious disease epidemiological models have been adopted to describe information propagation, largely due to several similarities between the spread of information and the transmission of infectious diseases in the population. For example, the susceptible-infected (SI) model \cite{1,2}, the susceptible-exposed-infected-recovered (SEIR) model \cite{3,4}, the susceptible-infected-recovered (SIR) model \cite{5,6} and the susceptible-infected-susceptible (SIS) \cite{7} model have all been utilized. In these traditional models, information propagation in social networks are modelled by stratifying users into three categories: heard rumor (ignorants), actively spreading rumor (spreaders), and no longer spreading rumor (stiflers) \cite{8}. Further extensions have been proposed to reflect novel propagation  mechanisms that distinguish from disease transmission, to make the extended models better take into account of specific factors that affect the information dissemination. Huang et al. \cite{9} constructed an improved rumor spreading SIR model, in which the impacts of rumor refuting by the affected enterprise, a microblogging opinion leader, and microblogging platform was analyzed. In consideration of eight influential factors related to rumor propagation, Zhang et al. \cite{10} developed a dynamic 8-state ignorance-carrier-spreader-advocate-removal (ICSAR) rumor propagation model. In 2018, Zhang et al. \cite{11} proposed a rumor spreading model with a view to event importance, event ambiguity, and public critical sense. Other studies concerned not only about the spread of rumors, but also about the spread of other kinds of information. In 2012, Zhang et al. \cite{12} proposed a time-varying hot topic propagation model on blog networks and bulletin board system sites, and in 2018 Jing et al. \cite{13} established an advertising spreading model applied to social networks. In other front, some studies have introduced new modules into the original model architecture. Liu et al. \cite{14} built a susceptible-antidotal-infected-removed (SAIR) model to characterize the super-spreading phenomenon in information propagation. In 2017, Wang et al. \cite{15} introduced a rumor spreading model called I2SR, in which spreaders are divided into those with a high rate of activity state and those with a low rate of activity state. Dong et al. \cite{16} introduced a varying total number of users and user deactivation rates to establish an improved SEIR rumor spreading model. In 2019, Zhang et al. \cite{17} established a susceptible-exposed-trusted-questioned-recovered (SETQR) model based on the traditional SEIR model and derived the stability of the SETQR model at the equilibrium point theoretically. 

In the process of information transmission, users play a decisive role. As such, several studies such as \cite{18,19,20} have incorporated the perspective of human psychology in modelling the information propagation. In these modeling frameworks, different attitudes of individuals, the evolutionary game and the level of user awareness are considered. However, in the field of information propagation, much less has been achieved in terms of incorporating the emotions of users into the dynamic models. We note that in 2019, Wang et al. \cite{21} proposed an emotion-based susceptible-infected-recovered (E-SIR) information propagation model. This model examined  how different emotions and diverse connections impact the information propagation and the possibility of information transmission between different users by dividing emotions of users into five categories: no emotion, anger, joy, disgust and sadness. They concluded that different emotions can impact on the scope of information propagation and the threshold of information explosion. 

To our best knowledge, no appropriate dynamic model framework has been developed and analyzed to examine the impact of the emotional choice of forwarding users on information propagation dynamics and emerging public sentiment. Here, we try to fill this gap by proposing an E-SFI model that considers personal emotions of users after exposing to related information with built in emotion from the forwarding users in the propagation chain. We develop and illustrate this framework, with model parameterized using the forwarding quantity that represents public attention to certain opinions about some public hot events.

\section{Model description}
\begin{figure}[htbp]
\centerline{\includegraphics[scale=0.6]{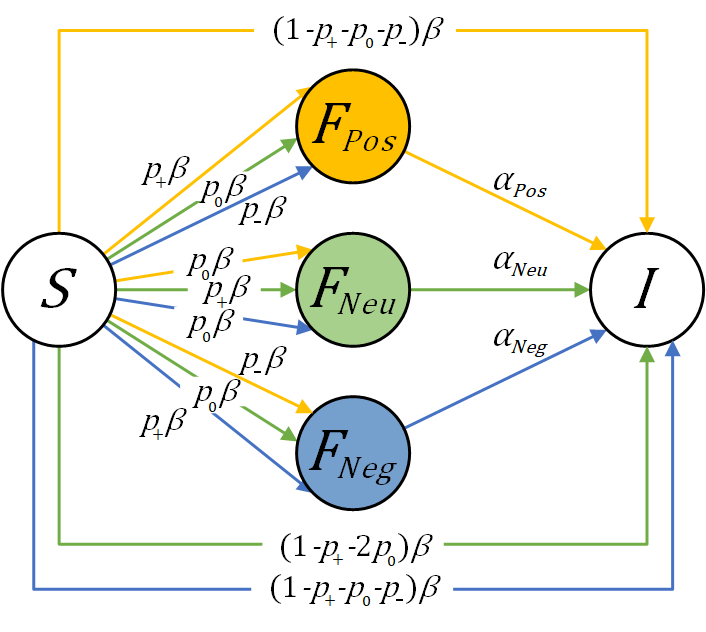}}
\caption{A schematic illustration of the information propagation along with emotions change, where susceptible users can be influenced by forwarding users who exposed to a piece of information with three possible emotions states: positive, neutral, negative, and then choose to forward the information with his/her own unique personal emotion.}
\label{fig2}
\end{figure}

Fig.~\ref{fig2} shows the structure of our E-SFI dynamic model proposed to investigate distinct public emotions triggered by particular events, when the emotion of original information is positive, or neutral, or negative.

In our model, we stratify the population $(N)$ into five states, and each user is in a unique state:
\begin{itemize}
    \item the susceptible state $(S)$, in which users are unaware of the information but susceptible to it;
    \item the positive forwarding state $(F_{Pos})$, the neutral forwarding state $(F_{Neu})$, and the negative forwarding state $(F_{Neg})$, in which users forward the information with positive emotion, neutral emotion and negative emotion, respectively. Users in these states have the ability to influence future susceptible users, also known as active individuals, meanwhile, these emotional differences come from the forwarding copywriting carried by users;
    \item the immune state $(I)$, in which users consist of two groups, one group of users who have  already forwarded the information with a specific emotion and will not forward it even if expose it again (so they are out of the active forwarding state), and the other group of users who are immune to the exposed information because they are subjectively not interested in the information.
\end{itemize}

And important parameters in the model are interpreted as follows:

\begin{itemize}
    \item  $\beta$: the average probability that a susceptible user can expose to the information, which can be influenced by the network structure;
    \item  $p_{+}$: the average probability that an individual in the susceptible state will forward the information with the same emotion as the previously contacted forwarding user;
    \item  $p_{0}$: the average probability that an individual in the susceptible state will forward the information with the slightly opposite emotion to the previously contacted forwarding use;
     \item  $p_-$: the average probability that an individual in the susceptible state will forward the information with the slightly opposite emotion to the previously contacted forwarding use;
     \item  $\alpha_{Pos}$: the average rate related to the behavior law of group users at which a user in the forwarding state with positive emotion $(F_{Pos})$  of information becomes inactive to forwarding; 
      \item  $\alpha_{Neu}$: the average rate related to the behavior law of group users at which a user in the forwarding state with neutral emotion $(F_{Neu})$ of information becomes inactive to forwarding;  
        \item  $\alpha_{Neg}$: the average rate related to the behavior law of group users at which a user in the forwarding state with negative emotion $(F_{Neg})$  of information becomes inactive to forwarding; 
\end{itemize}

We adopt basic structures in the SFI model \cite{22} and introduce the emotional choices of users into our established model. Our emotion-based susceptible-forwarding-immune (E-SFI) model takes the form below:

\begin{equation}\label{eq1}
\left\{
\begin{array}{l}
 
 dS(t)/dt=-\beta S(t)F_{Pos}(t)-\beta S(t)F_{Neu}(t)\\
\phantom{=\;\;\;\;\;\;\;\;\;\;\;}
 -\beta S(t)F_{Neg}(t)
\\dF_{Pos}(t)/dt=p_{+}\beta S(t)F_{Pos}(t)+p_{0}\beta S(t)F_{Neu}(t) \\ 
\phantom{=\;\;\;\;\;\;\;\;\;\;\;\;\;\;\;\;}
+p_{-}\beta S(t)F_{Neg}(t)-\alpha_{Pos}F_{Pos}(t)
\\ dF_{Neu}(t)/dt=p_{0}\beta S(t)F_{Pos}(t)+p_{+}\beta S(t)F_{Neu}(t) \\  
\phantom{=\;\;\;\;\;\;\;\;\;\;\;\;\;\;\;\;}
+p_{0}\beta S(t)F_{Neg}(t)-\alpha_{Neu}F_{Neu}(t)
\\ dF_{Neg}(t)/dt=p_{-}\beta S(t)F_{Pos}(t)+p_{0}\beta S(t)F_{Neu}(t)  \\ 
\phantom{=\;\;\;\;\;\;\;\;\;\;\;\;\;\;\;\;}
+p_{+}\beta S(t)F_{Neg}(t)-\alpha_{Neg}F_{Neg}(t)
\\ dI(t)/dt=(1-p_{+}-p_{-}-p_{0})\beta S(t)F_{Pos}(t) \\
\phantom{=\;\;\;\;\;\;\;\;\;\;}
+(1-p_{+}-2p_{0})\beta S(t)F_{Neu}(t)\\
\phantom{=\;\;\;\;\;\;\;\;\;\;}
+(1-p_{+}-p_{-}-p_{0})\beta S(t)F_{Neg}(t) \\ 
\phantom{=\;\;\;\;\;\;\;\;\;\;}
+\alpha_{Pos}F_{Pos}(t)+\alpha_{Neu}F_{Neu}(t)\\
\phantom{=\;\;\;\;\;\;\;\;\;\;}
+\alpha_{Neg}F_{Neg}(t)

\end{array}.
\right.
\end{equation}

An active forwarding user with specific emotion will contact an average number of $\beta N$ users per unit time, and the probability that a contacted user is a susceptible user is $S(t)/N$, so an active forwarding user will contact $\beta S(t)$ susceptible users. Since users who forward the information have three kinds of emotions, positive, neutral and negative, part of susceptible users will be influenced to forward information with personal emotions by forwarding users who have one of the above three emotions.

\begin{figure}[htbp]
\centerline{\includegraphics[scale=0.45]{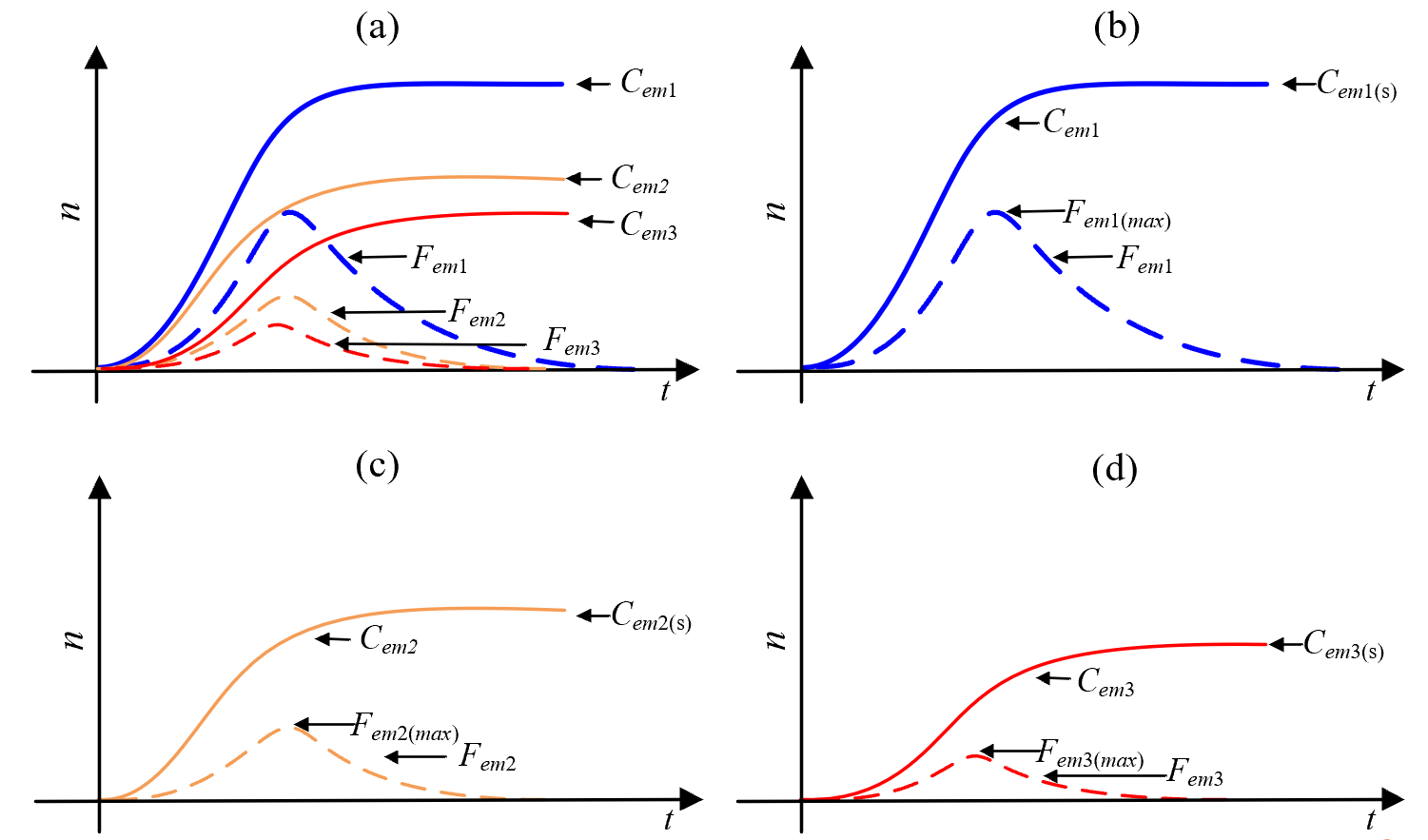}}
\caption{The temporal variation of the cumulative forwarding population and the instantaneous forwarding population with three kinds of emotions: (a) the temporal variation of two kinds of forwarding population with emotions in a hot event; (b) two indicators of emotion1; (c) two indicators of emotion2; (d) two indicators of emotion3.}
\label{fig3}
\end{figure}

In the process of single information propagation, the cumulative forwarding population and instantaneous forwarding population are two important indicators, and we can obtain these data from Chinese Sina-microblog. As shown in Fig.~\ref{fig3}, cumulative forwarding quantity with emotion1 $C_{em1}$ is greater than that with emotion2 $C_{em2}$, and cumulative forwarding quantity with emotion3 $C_{em3}$ is the least, which shows the scale of different public emotions caused by the information. According to that, in events with different emotional tones, emotion1, emotion2 and emotion3 are unfixed. For example, in an event where the emotional tone is positive, the emotion1 represents positive, however, in neutral or negative events, emotion1 represents neutral or negative respectively. The temporal variation of cumulative forwarding population (including $C_{em1}$, $C_{em2}$ and $C_{em3}$) is in the form of the increasing curve. The temporal variation of instantaneous forwarding population (including $F_{em1}$, $F_{em2}$ and $F_{em3}$) is in the form of the bell-shaped curve, which shows the explosive trend of different public emotions caused by the information. And we also can see that three cumulative emotional quantities of the information can be given by
\begin{equation}\label{eq2}
\begin{array}{l}
        C_{Pos}(t)=\int_{0}^{t}(p_{+}\beta S(t)F_{Pos}(t)+p_{0}\beta S(t)F_{Neu}(t)\\
\qquad\qquad+p_{-}\beta S(t)F_{Neg}(t))dt.
\end{array}
\end{equation}
\begin{equation}\label{eq3}
\begin{array}{l}
C_{Neu}(t)=\int_{0}^{t}(p_{0}\beta S(t)F_{Pos}(t)+p_{+}\beta S(t)F_{Neu}(t)\\
\qquad\qquad+p_{0}\beta S(t)F_{Neg}(t))dt.
\end{array}
\end{equation}
\begin{equation}\label{eq4}
\begin{array}{l}
 C_{Neg}(t)=\int_{0}^{t}(p_{-}\beta S(t)F_{Pos}(t)+p_{0}\beta S(t)F_{Neu}(t)\\
\qquad\qquad+p_{+}\beta S(t)F_{Neg}(t))dt.
\end{array}
\end{equation}

\section{Data fitting}
\noindent\textbf{Data description:}

In order to analyze the public sentiment, we selected the typical event with negative emotion and used the real data from Chinese Sina-Microblog. The accurate forwarding time data and forwarding copywriting for the forwarding record were collected from the Application Program Interface (API). We filtered the raw data to avoid the limitation of information stagnation caused by physiological needs.


\begin{table}[H]
\caption{Cumulative forwarding quantity with different emotions in the negative event-“An Australian Chinese woman who returns to Beijing refusing to quarantine goes out running”, which has a negative emotional keynote.}
\label{table2}
\scalebox{0.76}{
\begin{tabular}{c|c|c|c|c|c|c|c|c|c|c}
\hline
\textbf{t(30min)}  & \textbf{0}     & \textbf{1}     & \textbf{2}     & \textbf{3}     & \textbf{4}     & \textbf{5}     & \textbf{6}     & \textbf{7}     & \textbf{8}     & \textbf{9}    \\ \hline
$C_{Pos}$ & 68   & 165  & 262  & 355  & 454  & 545  & 634  & 757  & 866  & 963  \\ \hline
$C_{Neu}$ & 40   & 111  & 181  & 251  & 305  & 351  & 403  & 469  & 529  & 576  \\ \hline
$C_{Neg}$ & 265  & 639  & 950  & 1204 & 1470 & 1711 & 2015 & 2337 & 2629 & 2921 \\ \hline
\textbf{t(30min)}  & \textbf{10}    & \textbf{11}    & \textbf{12}    & \textbf{13}    & \textbf{14}    & \textbf{15}    & \textbf{16}    & \textbf{17}    & \textbf{18}    & \textbf{19}    \\ \hline
$C_{Pos}$ & 1035 & 1080 & 1095 & 1106 & 1112 & 1116 & 1121 & 1126 & 1130 & 1132 \\ \hline
$C_{Neu}$ & 613  & 627  & 631  & 634  & 635  & 638  & 644  & 646  & 650  & 654  \\ \hline
$C_{Neg}$ & 3147 & 3270 & 3323 & 3350 & 3374 & 3393 & 3421 & 3437 & 3461 & 3472 \\ \hline
\textbf{t(30min)}  & \textbf{20}    & \textbf{21}    & \textbf{22}    & \textbf{23}    & \textbf{24}    & \textbf{25}    & \textbf{26}    & \textbf{27}    & \textbf{…}     & \textbf{60}    \\ \hline
$C_{Pos}$ & 1132 & 1134 & 1135 & 1136 & 1136 & 1137 & 1137 & 1138 & …    & 1163 \\ \hline
$C_{Neu}$ & 658  & 658  & 658  & 658  & 658  & 658  & 658  & 658  & …    & 668  \\ \hline
$C_{Neg}$ & 3481 & 3484 & 3489 & 3491 & 3491 & 3493 & 3493 & 3496 & …    & 3573 \\ \hline
\end{tabular}}
\end{table}

The instantaneous forwarding quantity with different emotions can be distinguished by sentiment classification based on identifying words with polarity \cite{23,24} originating from forwarding copywriting, and then the cumulative forwarding quantity with different emotions can be obtained by adding the time series of instantaneous values within a certain time range. The beginning time is set to 0 and the sampling frequency is set to 30 minutes. Tables \ref{table2} lists part of the cumulative forwarding quantity with positive emotion ($C_{Pos}$), neutral emotion ($C_{Neu}$), and negative emotion ($C_{Neg}$) in the negative event, respectively. And the negative event, \#An Australian Chinese woman who returns to Beijing refusing to quarantine goes out running\#, has a negative keynote, so the negative forwarding is in the majority.

\noindent\textbf{Parameter estimation:}

In order to fit our model with real data from the Chinese Sina-microblog, we use the LS method to estimate the model parameters and the initial susceptible population. The parameter vector can be set as $\Theta=(\beta,p_+,p_0,p_-,\alpha_{Pos},\alpha_{Neu},\alpha_{Neg},S_0)$, and the corresponding numerical calculation based on the parameter vector for $C_{Pos}(t)$, $C_{Neu}(t)$ and $C_{Neg}(t)$ are denoted by $f_{C_{Pos}}(k,\Theta)$, $f_{C_{Neu}}(k,\Theta)$ and $f_{C_{Neg}}(k,\Theta)$, respectively. 

The LS error function
\begin{equation}\label{eq9}
\begin{split}
    LS=\sum\nolimits_{k=0}^T|f_{C_{Pos}}(k,\Theta)-C_{Posk}|^2\\
+\sum\nolimits_{k=0}^T|f_{C_{Neu}}(k,\Theta)-C_{Neuk}|^2\\
+\sum\nolimits_{k=0}^T|f_{C_{Pos}}(k,\Theta)-C_{Posk}|^2
\end{split}
\end{equation}
is used in our calculation, where $C_{Posk}$, $C_{Neuk}$ and $C_{Negk}$ denote the actual cumulative forwarding quantity with positive, neutral and negative emotion given in Tables \ref{table2}, and $k=0,1,2,…$ is the sampling time. We estimate the parameters of our E-SFI model with the data of the negative event.

\begin{figure}[htbp]
\centerline{\includegraphics[scale=0.25]{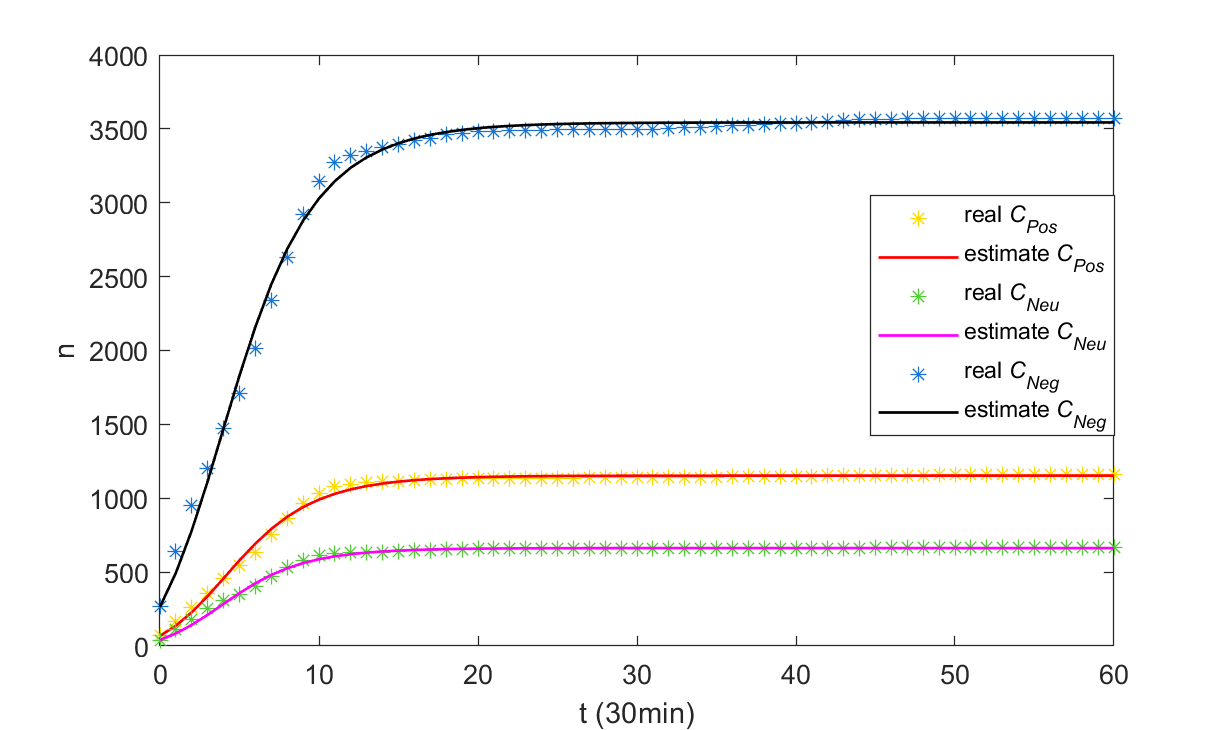}}
\caption{Numerical data fitting results of the negative event.}
\label{fig5}
\end{figure}

As shown in Fig. \ref{fig5}, we performed data fitting of the negative event on the real data given in Tables \ref{table2}, and the fitting curve is approximately consistent with the actual value. The orange star denotes the actual cumulative forwarding quantity with positive emotion, the green star denotes the actual cumulative forwarding quantity with neutral emotion, and the blue star denotes the actual cumulative forwarding quantity with negative emotion. The red line, purple line and black line denotes the estimated cumulative forwarding quantity with positive, neutral and negative emotion, respectively.

The average exposure rate $\beta$ is determined by the density of the propagation network, and the value of  $\beta$ is $1.2208\times10^{-4}$ in this event. And the probability of keeping emotions constant $p_+$ is $0.0263$, the probability of making emotion reverse $p_-$ is $9.7646\times10^{-4}$, and the probability of making emotion slightly reverse $p_0$ is $9.0446\times10^{-4}$, which indicates users are more willing to spread the information that they agree with, so as to support it. Meanwhile, the numerical simulation results show that $\alpha_{Pos}=0.4856$, $\alpha_{Neu}=0.5838$ and $\alpha_{Neg}=0.4373$ are similar. Stable cumulative forwarding quantity and maximum instantaneous forwarding quantity reveal the level and pattern of emotion-propagation. Event with the negative tone has the greatest stable cumulative and maximum instantaneous forwarding quantity with negative emotion $C_{Neg(s)}$ and $F_{Neg(max)}$. The results indicate that the most widespread emotion is determined by the emotional tone of the event.

\section{Discussion}

\begin{figure}[H]
\centerline{\includegraphics[scale=0.45]{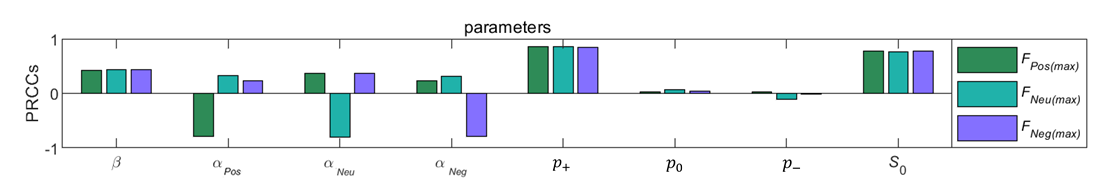}}
\caption{PRCC results with indices $F_{Pos(max)}$, $F_{Neu(max)}$, $F_{Neg(max)}$ of different parameters.}
\label{fig9}
\end{figure}

\begin{figure}[H]
\centerline{\includegraphics[scale=0.45]{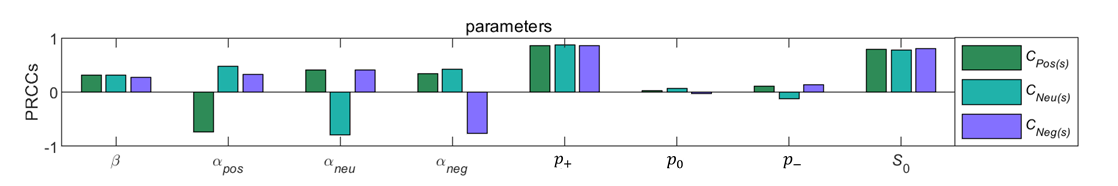}}
\caption{PRCC results with indices $C_{Pos(s)}$, $C_{Neu(s)}$, $C_{Neg(s)}$ of different parameters.}
\label{fig10}
\end{figure}

We use the partial rank correlation coefficients (PRCCs) \cite{25} to better discern the different parameters responsible for the E-SFI model, which is based on 1000 samples for various input parameters against the threshold condition to evaluate the sensitivity. In addition, 0.4 is chosen as the threshold, when the absolute value $|{\rm PRCC}|\geqslant0.4$, it is considered that there is a strong correlation between the index and the parameter, in other words, the index can be observably affected by changing this parameter, when the absolute value $0.2\leqslant|{\rm PRCC}|<0.4$, it means that this parameter has a median influence on the index, and when the absolute value $|{\rm PRCC}|<0.2$, the parameter works a weak correlation, that is, the larger the value, the greater the correlation between parameters and index. Fig. \ref{fig9}-\ref{fig10} give the PRCC results along with histograms, showing the effect of eight parameters ($\beta$, $\alpha_{Pos}$, $\alpha_{Neu}$, $\alpha_{Neg}$, $p_+$, $p_0$, $p_-$, $S_0$) on indices $F_{Neu(max)}$, $F_{Neg(max)}$, $C_{Pos(s)}$, $C_{Neu(s)}$, $C_{Neg(s)}$, in the E-SFI model respectively.

Fig. \ref{fig9}-\ref{fig10} show the influences of parameters on the maximum instantaneous forwarding quantity with different emotions $F_{Pos(max)}$, $F_{Neu(max)}$, $F_{Neg(max)}$ and stable cumulative forwarding quantity with different emotions $C_{Pos(s)}$, $C_{Neu(s)}$, $C_{Neg(s)}$. It can be seen from the comparative analysis that the influences of parameters on maximum instantaneous and stable cumulative forwarding quantity are very similar. Among that, the average exposure rate $\beta$, the probability of keeping emotion constant $p_+$ and the initial value $S_0$ have a positive effect on both forwarding quantity of information. As for average inactive rate with different emotions $\alpha_{Pos}$, $\alpha_{Neu}$ and $\alpha_{Neg}$, the maximum instantaneous forwarding quantity of certain emotion has a strong negative correlation with the same emotional average inactive rate and a relatively weak positive correlation with the other two emotional average inactive rates. For example, $\alpha_{Pos}$ has a strong negative effect on $F_{Pos(max)}$, while $\alpha_{Neu}$ and $\alpha_{Neg}$ have a correspondingly weak positive effect on it. The same conclusion applies to $F_{Neu(max)}$, $F_{Neg(max)}$ and the stable cumulative forwarding quantity with different emotions $C_{Pos(s)}$, $C_{Neu(s)}$, $C_{Neg(s)}$. The probability of making emotion slightly reverse $p_0$ and making emotion reverse $p_-$ have relatively weak impact on both forwarding quantity with different emotions from the results in these two figures.

From the above PRCC analysis, we can see that the average exposure rate $\beta$ and the probability of keeping emotion constant $p_+$ both have a significant influence on two considerable indicators, however influences of the probability of making emotion slightly reverse $p_0$ and the probability of making emotion reverse $p_-$ are hard to understand because the conclusion of sensitivity analysis with PRCC is stochastic to a certain extent. In this paper, $p_+$, $p_0$, $p_-$ are crucial to reflect the change of emotion, which is the focus of our research, for that, we need a more meticulous way to further analyze how do these  parameters affect the change of indices.

\begin{figure}[htbp]
\centerline{\includegraphics[scale=0.5]{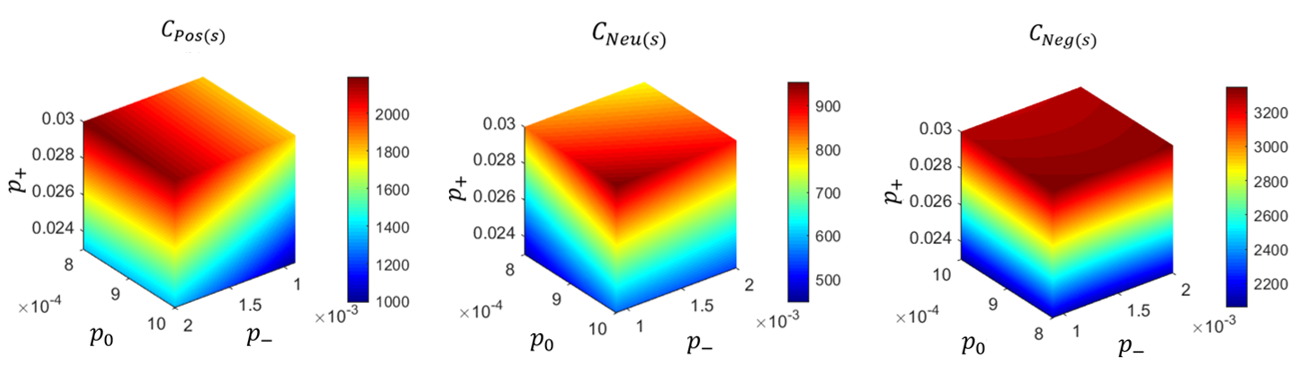}}
\caption{The comprehensive influence of multiple parameter variations on $C_{Pos(s)}$, $C_{Neu(s)}$ and $C_{Neg(s)}$ with the change of $p_-$, $p_0$ and $p_+$.}
\label{fig12}
\end{figure}

Fig. \ref{fig12} shows the effects of three parameters on values of the stable cumulative forwarding quantity with different emotions  $C_{Pos(s)}$, $C_{Neu(s)}$ and $C_{Neg(s)}$ respectively. The comparison and analysis of the  figure shows a specific pattern of conclusions between information with different emotions. The increase of $p_+$ is always going to obviously promote $C_{Pos(max)}$, $C_{Neu(max)}$, $C_{Neg(max)}$, however, the impacts are different. $p_+$ has a greater influence on stable cumulative forwarding quantity which with the same emotion as the original information, and is $C_{Neg(max)}$ in this example getting from the value of the figure. With the increase of $p_+$ and $p_-$, $C_{Pos(max)}$ increases and $p_0$ has almost no influence on it. Whereas with the increase of $p_+$ and $p_0$, with the decrease of $p_-$, $C_{Neu(max)}$ increases. As for $C_{Neg(max)}$, the change of both $p_0$ and $p_-$ has a weak influence compared to $p_+$ on it, so with the increase of $p_+$, $C_{Neg(max)}$ can get apparently increase. 

We have performed the data fitting based on real data obtained from Chinese Sina-microblog to verify the effectiveness of the E-SFI model. Based on our analytic study and numerical simulations, we obtain  some interesting insights into the public sentiment formation and information with emotion propagation. Namely, we note that the average inactive rate is largely related to the behavior rules of group users, which is difficult to be changed and interposed in a short time theoretically. The initial susceptible population, average exposure rate and three kinds of forwarding probabilities can be adjusted so as to make the expected emotion to spread or dissipate. 

More specifically, if we want the information to be disseminated on a large scale, we can achieve it by selecting the information diffusion source with a higher network density at the beginning of the information propagation to obtain a larger average exposure rate $\beta$. By attracting opinion leaders, who have a large number of fans and appealing power to forward the original information, we will recruit more groups interested in the event, thereby increasing the initial value $S_0$, and ultimately promoting information dissemination. In addition, when the original post is in the desired positive or neutral state, we should work with opinion leaders to forward the original post with the same emotion, thus increasing the probability of keeping emotion constant $p_+$. If the original post is in the undesired negative state, a best way is to work with opinion leaders to forward the original post with one of the other two emotions, so as to raise the probability of making emotion slightly reverse $p_0$ or the probability of making emotion reverse $p_-$. Although the emotional tone of the original information determines the mainstream emotion in forwarding, it is still feasible to increase the forwarding quantity of the information with altered emotions by changing slightly opposite or opposite-forwarding probabilities. On the contrary, reducing the attention of the original information and suppress public sentiment is possible by preventing the opinion leaders from forwarding the original information by reducing the initial value $S_0$, changing the probability of keeping emotion constant $p_+$, the probability of making emotion slightly reverse $p_0$ and the probability of making emotion reverse $p_-$, and reducing the average exposure rate $\beta$ by truncating the key nodes.

\section{Conclusion}
 Our emotion-based susceptible-forwarding-immune (E-SFI) dynamic model considers the general behavior that when a user spreads the emotional information, the user will carry the same, slightly opposite or opposite personal emotion along with the information and emotion received/exposed. We have performed numerical simulations based on the real data related to three events with different emotional tones to verify the effectiveness of our model. We concluded that normally the amount of forwarding which is consistent with the emotional tone of the original information is far more than that of other emotions, but the other emotions coming along with the information through this process should not be ignored. We performed sensitivity analyses on parameters and found that the probability of keeping emotions constant $p_+$, the probability of making emotion slightly reverse $p_0$ and the probability of making emotion reverse $p_-$ are all significant in terms of guiding the public sentiment. We hope our E-SFI dynamic model fills the theoretical gap and provides support to construct effective communication strategies in the online network environment.

\section*{Acknowledgments}
The work was supported by the National Natural Science Foundation of China (Grant numbers: 61801440), the Natural Science and Engineering Research Council of Canada, the Canada Research Chair Program (JWu), the Fundamental Research Funds for the Central Universities, State Key Laboratory of Media Convergence and Communication, Communication University of China and the High-quality and Cutting-edge Disciplines Construction Project for Universities in Beijing (Internet Information, Communication University of China).




\begin{thebibliography}{1}


\bibitem{1}
F. Chen, “A susceptible-infected epidemic model with voluntary vaccinations,” Journal of mathematical biology, vol. 53, pp. 253-272, 2006.
\bibitem{2}
Z. Lu, S. Gao, and L. Chen, “Analysis of an SI epidemic model with nonlinear transmission and stage structure,” Acta Mathematica Scientia, vol. 23, pp. 440-446, 2003.
\bibitem{3}
D. Xiao and S. Ruan, “Global analysis of an epidemic model with nonmonotone incidence rate,” Mathematical biosciences, vol. 208, pp. 419-429, 2007.
\bibitem{4}
M. Li and J. S. Muldowney, “Global stability for the SEIR model in epidemiology,” Mathematical biosciences, vol. 125, pp. 155-164, 1995.
\bibitem{5}
W. O. Kermack and A. G. McKendrick, “Contributions to the mathematical theory of epidemics—I,” Bulletin of mathematical biology, vol. 53, pp. 33-55, 1991.
\bibitem{6}
L. Stone, B. Shulgin, and Z. Agur, “Theoretical examination of the pulse vaccination policy in the SIR epidemic model,” Mathematical and computer modelling, vol. 31, pp. 207-215, 2000.
\bibitem{7}
C. Xia, S. Sun, F. Rao, and J. Sun, “SIS model of epidemic spreading on dynamical networks with community,” Frontiers of Computer Science in China, vol. 3, pp. 361-365, 2009.
\bibitem{8}
D. J. Daley and D. G. Kendall, “Epidemics and rumours,” Nature, vol. 204, pp. 1118-1118, 1964.
\bibitem{9}
J. Huang and Q. Su, “A rumor spreading model based on user browsing behavior analysis in microblog,” IEEE 2013 10th International Conference on Service Systems and Service Management, pp. 170-173, 2013.
\bibitem{10}
N. Zhang, H. Huang, B. Su, J. Zhao, and B. Zhang, “Dynamic 8-state ICSAR rumor propagation model considering official rumor refutation,” Physica A: Statistical Mechanics and Its Applications, vol. 415, pp. 333-346, 2014.
\bibitem{11}
Y. Zhang, J. Xu, and Y. Wu, “A fuzzy rumor spreading model based on transmission capacity,” International Journal of Modern Physics C, vol. 29, pp. 1850012, 2018.
\bibitem{12}
B. Zhang, X. Guan, M. J. Khan, and Y. Zhou, “A time-varying propagation model of hot topic on BBS sites and Blog networks,” Information Sciences, vol. 187, pp. 15-32, 2012.
\bibitem{13}
Y. Jing, P. Liu, X. Tang, and W. Liu, “Improved SIR Advertising Spreading Model and Its Effectiveness in Social Network,” Procedia Computer Science, vol. 129, pp. 215-218, 2018.

\bibitem{14}
Y. Liu, B. Wang, B. Wu, S. Shang, Y. Zhang, and C. Shi, “Characterizing super-spreading in microblog: An epidemic-based information propagation model,” Physica A: Statistical Mechanics and its Applications, vol. 463, pp. 202-218, 2016.
\bibitem{15}
L. Wang, N. Song, C. Ma, and B. He, “Rumor spreading model considering the activity of spreaders in the homogeneous network,” Physica A: Statistical Mechanics and its Applications, vol. 468, pp. 855-865, 2017.
\bibitem{16}
S. Dong, Y. Deng, and Y. Huang, “SEIR model of rumor spreading in online social network with varying total population size,” Communications in Theoretical Physics, vol. 68, pp. 545-552, 2017.
\bibitem{17}
Y. Zhang and Z. Chen, “SETQR Propagation Model for Social Networks,” IEEE Access, vol. 7, pp. 127533-127543, 2019.
its Applications, vol. 535, pp. 122236, 2019.
\bibitem{18}
Y. Hu, Q. Pan, W. Hou, and M. He, “Rumor spreading model with the different attitudes towards rumors,” Physica A: Statistical Mechanics and its Applications, vol. 502, pp. 331-344, 2018.
\bibitem{19}
Y. Xiao, D. Chen, S. Wei, Q. Li, H. Wang, et al, “Rumor propagation dynamic model based on evolutionary game and anti-rumor,” Nonlinear Dynamics, vol. 95, pp. 523-539, 2019.
\bibitem{20}
C. Sang and S. Liao, “Modeling and simulation of information dissemination model considering user’s awareness behavior in mobile social networks,” Physica A: Statistical Mechanics and its Applications, vol. 537, pp. 122639, 2020.
\bibitem{21}
T. Wang, M. Hu, and L. Kou, “A Information Propagation Model Based on Various Emotions and Heterogeneous Mean Field in Social Networks,” 12th EAI International Conference on Mobile Multimedia Communications, European Alliance for Innovation, 2019.
\bibitem{22}
F. Yin, X. Shao, and J. Wu, “Nearcasting forwarding behaviors and information propagation in Chinese Sina-Microblog,” Mathematical Biosciences and Engineering, vol. 16, pp. 5380-5394, 2019.
\bibitem{23}
Y. Cai, K. Yang, D. Huang, Z. Zhou, L. Xue, et al, “A hybrid model for opinion mining based on domain sentiment dictionary,” International Journal of Machine Learning and Cybernetics, pp. 1-12, 2019.
\bibitem{24}
K. Jia and Z. Li, “Chinese Micro-Blog Sentiment Classification Based on Emotion Dictionary and Semantic Rules,” IEEE 2020 International Conference on Computer Information and Big Data Applications, pp. 309-312, 2020.
\bibitem{25}
U. L. Abbas, R. M. Anderson, and J. W. Mellors, “Potential Impact of Antiretroviral Chemoprophylaxis on HIV-1 Transmission in Resource-Limited Settings,” PLOS ONE, vol. 2, 2007.
\end{thebibliography}
%

\end{document}